\documentclass[aps,prl,twocolumn,superscriptaddress,groupedaddress, floatfix]{revtex4}
\usepackage[utf8]{inputenc}
\usepackage{amsmath}
\usepackage[sectionbib]{chapterbib}
\usepackage{graphicx}
\usepackage[usenames, dvipsnames]{color}
\usepackage{float}
\usepackage{array}
\usepackage{color}
\graphicspath{{Figures/}}

\newcommand{\ket}[1]{| #1 \rangle}

\newcommand{\MHz}{\mathrm{MHz}}
\newcommand{\GHz}{\mathrm{GHz}}

\newcommand{\nm}{\mathrm{nm}}
\newcommand{\mK}{\mathrm{mK}}

\newcommand{\us}{\mathrm{\mu}\mathrm{s}}

\newcommand{\ms}{\mathrm{ms}}

\newcommand{\n}{n_\mathrm{g}}
\newcommand{\df}{\delta f_{01} (\n)}

\newcommand{\Ej}{E_\mathrm{J}}
\newcommand{\Ec}{E_\mathrm{C}}
\newcommand{\sgam}[1]{\Gamma_{#1}^{\mathrm{eo}}}
\newcommand{\ngam}[1]{\Gamma_{#1}^{\mathrm{ee}}}
\newcommand{\F}{\mathcal{F}}
\newcommand{\xqpn}{x_\mathrm{qp}^0}

\newcommand{\nlb}{\nolinebreak}
\newcommand{\pp}[1]{\langle PP'\rangle_{#1}}
\newcommand{\tng}{\tau(\n)}

\begin{document}

\widetext

\title{Hot nonequilibrium quasiparticles in transmon qubits}

\author{K.~Serniak}
\email{kyle.serniak@yale.edu}
\affiliation{Department of Applied Physics, Yale University, New Haven, CT 06520, USA}
\author{M.~Hays}
\affiliation{Department of Applied Physics, Yale University, New Haven, CT 06520, USA}
\author{G.~de~Lange}
\affiliation{Department of Applied Physics, Yale University, New Haven, CT 06520, USA}
\affiliation{QuTech and Kavli Institute of Nanoscience, Delft University of Technology, 2600 GA Delft, The Netherlands}
\author{S.~Diamond}
\affiliation{Department of Applied Physics, Yale University, New Haven, CT 06520, USA}
\author{S.~Shankar}
\affiliation{Department of Applied Physics, Yale University, New Haven, CT 06520, USA}
\author{L.~D.~Burkhart}
\affiliation{Department of Applied Physics, Yale University, New Haven, CT 06520, USA}
\author{L.~Frunzio}
\affiliation{Department of Applied Physics, Yale University, New Haven, CT 06520, USA}
\author{M.~Houzet}
\affiliation{Univ. Grenoble Alpes, CEA, INAC-Pheliqs, F-38000 Grenoble, France }
\author{M.~H.~Devoret}
\email{michel.devoret@yale.edu}
\affiliation{Department of Applied Physics, Yale University, New Haven, CT 06520, USA}

\date{\today}
\begin{abstract}
Nonequilibrium quasiparticle excitations degrade the performance of a variety of superconducting circuits.
Understanding the energy distribution of these quasiparticles will yield insight into their generation mechanisms, the limitations they impose on superconducting devices, and how to efficiently mitigate quasiparticle-induced qubit decoherence. 
To probe this energy distribution, we systematically correlate qubit relaxation and excitation with charge-parity switches in an offset-charge-sensitive transmon qubit, and find that quasiparticle-induced excitation events are the dominant mechanism behind the residual excited-state population in our samples. 
By itself, the observed quasiparticle distribution would limit $T_1$ to $\approx\nlb200~\us$, which indicates that quasiparticle loss in our devices is on equal footing with all other loss mechanisms.
Furthermore, the measured rate of quasiparticle-induced excitation events is greater than that of relaxation events, which signifies that the quasiparticles are more energetic than would be predicted from a thermal distribution describing their apparent density.
\end{abstract}

\maketitle
\chapter{}
\indent The adverse effects of nonequilibrium quasiparticles (QPs) ubiquitous in aluminum superconducting devices have been recognized in a wide variety of systems, including Josephson junction (JJ) based superconducting qubits~\cite{Aumentado2004,Lutchyn2005,Shaw2008,Martinis2009,Lenander2011,Catelani2011,Sun2012,Wenner2013,Riste2013,Pop2014,Vool2014,Bal2015,Riwar2016}, kinetic-inductance~\cite{Day2003,Monfardini2012,Grunhaupt2018} and quantum-capacitance~\cite{Stone2012} detectors, devices for current metrology~\cite{Pekola2008}, Andreev qubits~\cite{Olivares2014,Janvier2015,Hays2017}, and proposed Majorana qubits~\cite{Higginbotham2015,Albrecht2017}.
While recent efforts to reduce the density of QPs in superconducting qubits have shown some improvement in the relaxation times of devices limited by QP-induced loss~\cite{Nsanzineza2014, Vool2014, Wang2014, Gustavsson2016}, understanding the energy distribution of nonequilibrium QPs may shed light on their source and further help to mitigate their effects. Furthermore, it has been suggested that ``hot" nonequilibrium QPs may be responsible for the residual excited state population seen in superconducting qubits at low temperatures~\cite{Wenner2013,DeVisser2014a,Jin2015}, though this has yet to be confirmed directly.\\
\indent In this letter, we report signatures of hot nonequilibrium QPs observed in the correlations between qubit transitions and QP-tunneling events.
An offset-charge sensitive transmon qubit was used to directly detect switches in the charge-parity of the transmon islands associated with individual QPs tunneling across the JJ~\cite{Riste2013}.
We correlated these charge-parity switches with transitions between the ground and first-excited states of the transmon, and found that QP tunneling accounts for $\approx\nlb 30\%$ of all qubit relaxation events and $\approx\nlb 90\%$ of excitation events.
The measured ratio of the QP-induced excitation and relaxation rates is greater than one, which is at odds with a thermal distribution accounting for their estimated density, defining what we refer to as a ``hot" energy distribution of tunneling QPs. 
These results confirm previous suspicions that nonequilibrium QPs are responsible for the residual excited state population in transmon qubits~\cite{Wenner2013,DeVisser2014a,Jin2015}, and emphasize the need for further understanding of QP-induced loss.\\
\begin{figure}[b]
	\centering
    \includegraphics[width=\columnwidth]{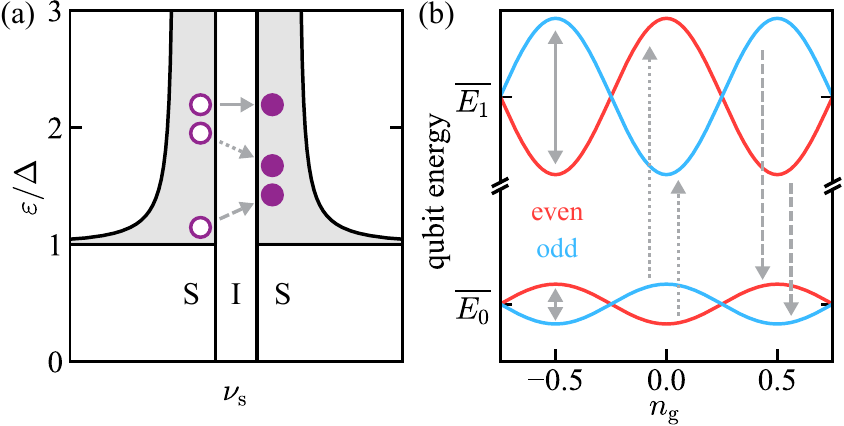} 
	\caption{
    QP-induced transitions in transmon qubits.
    (a) Density of states $\nu_\mathrm{s}$ versus the reduced energy $\varepsilon/\Delta$ in the leads of a superconductor-insulator-superconductor (SIS) JJ, in the excitation representation.
    Grey arrows represent tunneling processes of QPs, shown as purple dots.
    Dashed, dotted, and solid lines correspond to relaxation, excitation, and inter-band transitions of the qubit, respectively, with associated inelastic QP scattering.
    (b) The two lowest energy levels of an offset-charge-sensitive transmon qubit (vertical axis not to scale) as a function of offset-charge $\n$, in units of $2e$. These levels are shifted depending on the charge parity (even or odd) of the qubit, and $\overline{E_0}$ and $\overline{E_1}$ are time-averaged energies of the ground and first-excited states, respectively, assuming ergodic fluctuations of $\n$ and/or charge parity.
    Arrows correspond to those in (a).
    }
\end{figure}
Ideally, QPs in superconducting devices would be in thermal equilibrium with their thermal anchor~($T\approx\nlb 20~\mK$ for dilution refrigerators), and their spontaneous generation would be exponentially suppressed by the superconducting gap~$\Delta$. 
However, there is an observed fraction of broken Cooper pairs~$\xqpn\approx\nolinebreak10^{-8}\text{-}10^{-6}$~\cite{Aumentado2004,Segall2004,Martinis2009,Shaw2008,DeVisser2014,Wang2014,Vool2014,DeVisser2014a,Taupin2016} which is orders of magnitude greater than would be predicted in thermal equilibrium.
In a transmon~\cite{Koch2007}, QP tunneling across the JJ will always change the excess charge on the islands by $1\mathrm{e}$, switching the charge parity of the junction electrodes between ``even" and ``odd"~\cite{Lutchyn2005}.  
Tunneling QPs couple to the phase across the JJ~\cite{Martinis2009,Catelani2011}, and consequently can induce qubit transitions~[Fig.~1].
If the QPs were in thermal equilibrium, the values of $\xqpn$ quoted above would correspond to an effective QP temperature of~$130\text{-}190~\mK$. 
Under this assumption, QP-induced relaxation of the qubit should vastly outweigh QP-induced excitation. As we will show, this is not observed in our devices, indicating that this effective temperature does not adequately describe the QP energy distribution.\\
\indent To directly probe the interaction between nonequilibrium QPs and a transmon qubit, we slightly relax the transmon-defining condition that the Josephson coupling energy~$\Ej$ is much greater than the charging energy~$\Ec$~\cite{Sun2012}. 
In this regime, the ground-to-excited-state-transition frequency $f_{01} =\nlb(E_1 - E_0)/h$ has a measurable dependence on charge parity, switching between $\overline{f_{01}}\pm\delta f_{01}$ when a QP tunnels across the JJ~(the qubit energies switch between the blue and red lines in~Fig.~1b)~\cite{Sun2012,Riste2013}.
The deviation~$\delta f_{01}$ is a sinusoidal function of the dimensionless offset-charge~$\n$, which undergoes temporal fluctuations due to reconfiguration of mobile charges in the environment.
Because $h\df\ll\nlb k_\mathrm{B} T$, QP tunneling dynamics will not depend strongly on $\n$. 
The authors of Ref.~\cite{Riste2013} took advantage of this frequency splitting to track~$\n$, map the charge parity onto the state of a transmon, and correlate qubit relaxation with parity switches~\cite{Catelani2014}.
Extending their experiment, we extract not only the QP-induced relaxation rate, but also the  QP-induced excitation rate by detailed modeling of the correlations between charge-parity switches and qubit transitions. \\
\indent We focus below on a single transmon qubit with average frequency $\overline{f_{01}}=\nlb4.400~\GHz$ and $\Ej/\Ec=\nlb23$, corresponding to a maximum even-odd splitting $2\delta f_\mathrm{01}(0)=\nlb3.18~\MHz$. 
The average measured relaxation time~$T_1=\nolinebreak95~\us$ is on par with state-of-the-art transmons, and the equilibrium ground state population~$\mathcal{P}_\mathrm{0}^\mathrm{eq}=\nlb 0.74$ corresponds to an effective qubit temperature of~$160$~mK. 
Data from a second sample with similar parameters is discussed in the Supplemental Material~\cite{SOM}. 
Chips were mounted in an Al 3D rectangular readout cavity~\cite{Paik2011} and anchored to the mixing chamber of a cryogen-free dilution refrigerator at~$20~\mathrm{mK}$.\\
\begin{figure}
	\centering
	\includegraphics[width=\columnwidth]{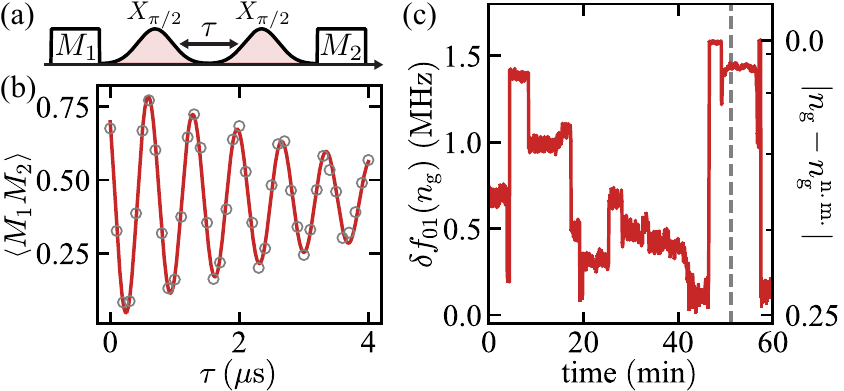} 
	\caption{Monitoring slow fluctuations of~$\df$.
    (a) Depiction of the Ramsey sequence. 
    High-fidelity qubit measurements $M_1$ and $M_2$ have thresholded outcome~$0$~or~$1$, corresponding to the ground and first-excited states of the qubit, respectively. 
    (b) Ramsey fringes of~$\langle M_1 M_2 \rangle$ oscillate at $\df$, which is measured every~$\sim 4~\mathrm{s}$~(c). The grey dashed line marks the frequency fit from (b). 
  	The right-side y-axis shows the conversion from $\df$ to $\n$, where $\n^\mathrm{n.m.}$ is the value of $\n$ corresponding to the nearest maximum of $\df$.
    }
\end{figure}
The slow background fluctuations of $\n$ were tracked by monitoring $\df$ using the Ramsey sequence depicted in Fig. 2(a).
The carrier frequency of the Gaussian $\pi/2$-pulses is chosen to be~$\overline{f_{01}}$, which is symmetrically detuned from the even and odd charge-parity states at all values of~$\n$.
This ensures that the phase evolution of even- and odd-parity states on the equator of the Bloch sphere will interfere constructively, resulting in Ramsey fringes~[Fig.~2(b)] characterized by a single oscillation frequency~$\df$ and a decay constant~$T_2$ that is insensitive to fast charge-parity switches. 
Repeated Ramsey experiments~[Fig.~2(c)] show that $\n$ fluctuates on a timescale of minutes, which is long enough to perform experiments that rely on prior knowledge of $\df$.\\
\indent Using a similar pulse sequence~[Fig.~3(a)], we map the charge parity of the transmon onto the qubit state~\cite{Riste2013}.
Two~$\pi/2$-pulses, now about orthogonal axes, are separated by a delay~$\tau(\n)=\nlb 1/4\df$, which constitutes an effective $\pi$-pulse conditioned on charge parity~($\pi_\mathrm{e,o}$).
This charge-parity-mapping operation only discerns between transition frequencies greater-than or less-than $\overline{f_{01}}$, and we refer to these as ``even" and ``odd'' charge-parity states, respectively, despite the inability to measure absolute parity. The relative phase of the $\pi/2$-pulses controls whether the $\pi_\mathrm{e,o}$ sequence is conditioned on even or odd charge parity.
The charge parity \mbox{$P =\nlb(2 M_1-1)(2M_2 - 1)$} is calculated in post-processing. 
To observe QP-tunneling events in real time, we repeated the charge-parity-mapping sequence every~$\Delta t_\mathrm{exp}=\nlb 10~\us$ for $\sim\nlb600~\ms$~[Fig.~3(b)]. 
The power spectral density~$S_{PP}$ of these parity fluctuations was averaged over~$20$ independent charge-parity jump traces~[Fig.~3]. 
$S_{PP}$ was fit to the characteristic Lorentzian of a random telegraph signal, from which a parity-switching timescale $T_P=\nlb77\pm\nlb1~\us$ and mapping fidelity $\F=\nlb0.91\pm\nlb0.01$ were obtained~\cite{SOM}. 
Each jump trace was acquired after confirming that $\df>1~\MHz$ by the monitoring of $\n$ described above.
This conditioning was introduced to increase the fidelity~$\F$ of the parity mapping, as $\df$ is less sensitive to fluctuations in $\n$ at near-maximum $\df$; also, the qubit is less likely to dephase during the correspondingly shorter~$\tau(\n)$.\\
\begin{figure}
	\centering
	\includegraphics[width=\columnwidth]{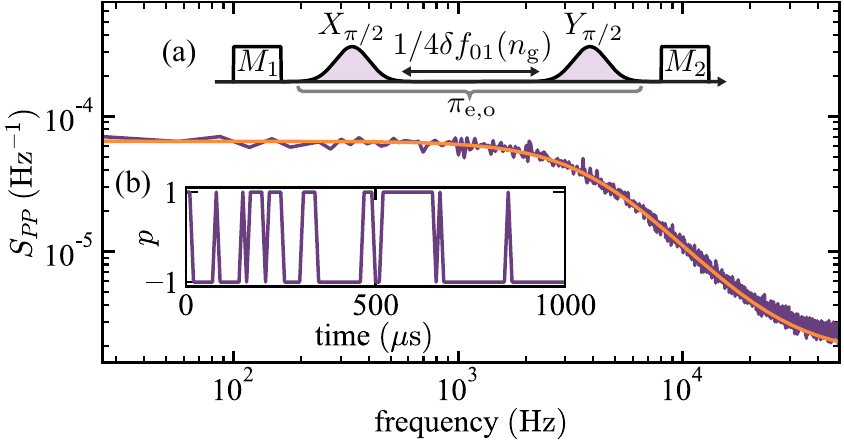} 
	\caption{Detecting fast charge-parity switches in an offset-charge-sensitive transmon qubit.
	(a) Charge-parity mapping pulse sequence, which results in an effective charge-parity-conditioned $\pi$-pulse,~$\pi_\mathrm{e,o}$.
    Inset (b): A~$1~\ms$ snapshot of a~\mbox{$\sim 600~\ms$} long charge-parity jump trace.
    Main: Power-spectrum of charge-parity fluctuations,  with a Lorentzian fit (orange) corresponding to \mbox{$T_P=77\pm1~\us$}.
    }
\end{figure}
\begin{figure*}
	\centering
	\includegraphics[width=\textwidth]{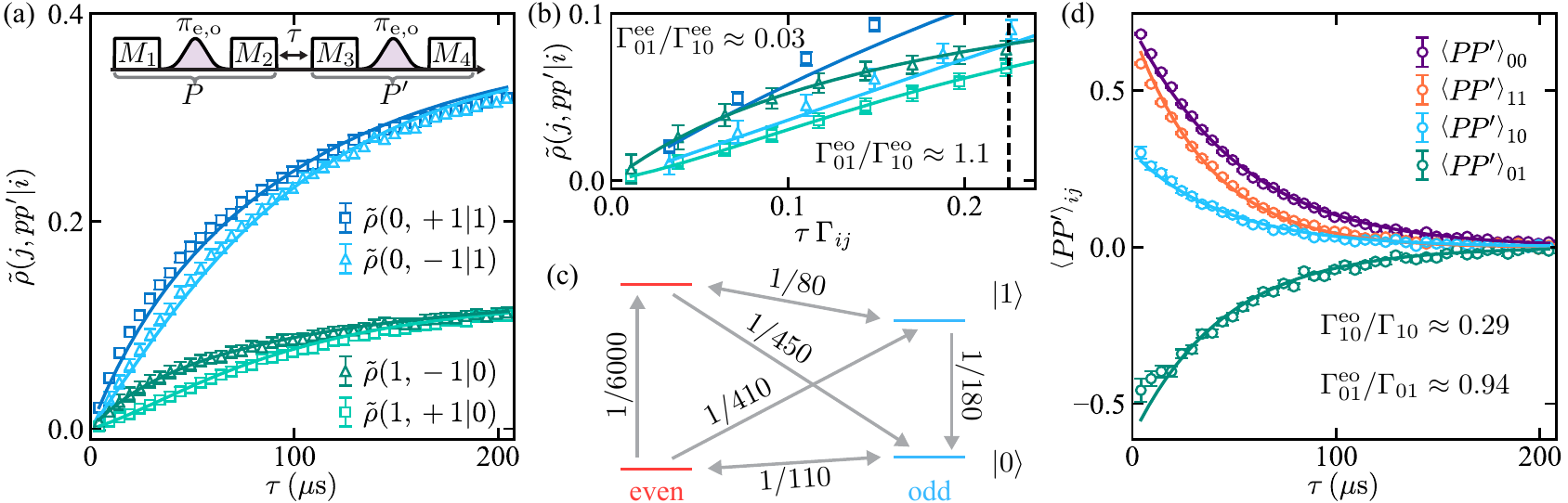} 
	\caption[width=\textwidth]{Correlating charge-parity switches with qubit transitions.
    (a) Inset: Pulse sequence depicting the charge-parity correlation measurement. The charge-parity conditioning of the state-mapping sequence is varied between measurements to balance mapping-dependent errors.
    Main: Conditioned probabilities $\tilde{\rho}(j,pp'|i)(\tau)$ with and without a charge-parity switch ($pp'=+1$ or $-1$, respectively).
    The relative amplitudes of curves with and without parity switches (triangles and squares, respectively) indicate the likelihood that those transitions were correlated with quasiparticle-tunneling events.
    Theory lines are obtained from a least-squares fit to the master equation described in the main text.
    (b) Probabilities plotted in (a) after rescaling $\tau$ by $\Gamma_{ij}$, the overall decay rate governing each curve at large $\tau$. The crossing of curves with $pp'=-1$ (black-dashed line) indicates a negative effective temperature of the quasiparticle bath.
    (c) Transition rates extracted from the master equation, in units of $\us^{-1}$. Note that rates are invariant under exchange of even and odd charge-parity states.
    (d) Charge-parity autocorrelation function $\langle PP'\rangle$ conditioned on the outcomes $m_2=i$ and $m_3=j$.
    }   
\end{figure*}
The fact that $T_P\approx T_1$ hints at the possibility that our transmon may be limited by QP-induced dissipation. 
Following Ref.~\cite{Catelani2014}, the total relaxation rate $\Gamma_{10}$ can be decomposed into the sum of two contributions: the rate of relaxation accompanied by a charge-parity switch ($\sgam{10}$), which we attribute solely to QP-induced loss, and the rate of relaxation from charge-parity-conserving mechanisms ($\ngam{10}$), such as dielectric loss.
As there is no preferred parity, these transition rates are symmetric under exchange of even and odd ($\Gamma_{ij}^\mathrm{eo} = \Gamma_{ij}^\mathrm{oe}$ and $\Gamma_{ij}^\mathrm{ee} = \Gamma_{ij}^\mathrm{oo}$). 
Similarly to the total relaxation rate, the total excitation rate is given by $\Gamma_{01}=\sgam{01}+\ngam{01}$.
We resolve these distinct contributions by concatenating two parity-mapping sequences (outcomes $p$ and $p'$) separated by a variable delay $\tau$~[Fig.~4(a), inset].
This measurement determines both the charge parity and qubit state before and after $\tau$, which allows us to correlate qubit transitions with QP tunneling events.
From our data, we compute $\tilde{\rho}(j,pp'|i)(\tau)$: the probability of measuring outcome $m_3=\nlb j$ after a delay $\tau$ given that $m_2 =\nlb i$, with or without a parity switch~($pp'=-1$ or $+1$, respectively).
To model these quantities, we employ a master equation describing the flow of probability between different system states
\begin{equation}
\begin{split}
\dot{\rho_i^\alpha}=&-(\Gamma_{i\bar{i}}^{\alpha\bar{\alpha}}+\Gamma_{ii}^{\alpha\bar{\alpha}}+\Gamma_{i\bar{i}}^{\alpha\alpha})\rho_i^\alpha \\&+ \Gamma_{\bar{i}i}^{\bar{\alpha}\alpha} \rho_{\bar{i}}^{\bar{\alpha}}+\Gamma_{ii}^{\bar{\alpha}\alpha} \rho_{i}^{\bar{\alpha}}
+\Gamma_{\bar{i}i}^{\alpha\alpha} \rho_{\bar{i}}^{\alpha},
\end{split}
\end{equation}
where $\rho_i^\alpha$ is the probability of finding the system in qubit state $i$ and charge parity $\alpha$, and $\overline{i}$ is read as ``not~$i$."
We evolve the above model numerically with initial conditions determined by $M_2$ and $P$, and fit all eight conditional probabilities $\tilde{\rho}(j,pp'|i)(\tau)$, a subset of which are shown in~Fig.~4(a,~b).\\
\indent In addition, we calculate the charge-parity autocorrelation function $\langle PP'\rangle_{ij}(\tau)$, again conditioned on $m_2=\nlb i$ and $m_3=\nlb j$, respectively~[Fig.~4(d)], and fit to functions of the form~\cite{SOM}
\begin{equation}
\langle PP'\rangle_{ij}(\tau)=\rho^{\alpha}_{i}(0)\left(\frac{\rho^\alpha_j(\tau)-\rho^{\bar{\alpha}}_j(\tau)}{\rho^\alpha_j(\tau)+\rho^{\bar{\alpha}}_j(\tau)}\right).
\end{equation}
The maximum correlation $\langle PP'\rangle_{ii}(0)$ is limited by the fidelity of the correlation measurement, and  qualitatively, the deviation of $\langle PP'\rangle_{ij}(0)$ from this maximum amplitude is related to the ratio $\sgam{ij}/\Gamma_{ij}$~[Fig. 4(d)].\\
\indent Equations (1) and (2) do not account for any measurement infidelities, which can skew the observed correlations.
These include parity- and qubit-state-dependent errors, such as spontaneous qubit transitions during the parity-mapping sequence, as well as global errors such as pulse infidelity due to uncertainty in~$\df$. 
We stress that proper modeling of these errors is necessary to accurately extract the conditional rates. 
Taking into account these considerations, we fit all eight permutations of $\tilde{\rho}(j,pp'|i)(\tau)$ and the four $\langle PP'\rangle_{ii}(\tau)$ curves simultaneously to the master equation model~(solid lines in Fig.~4). For more details on the model and fit, see the Supplemental Material~\cite{SOM}.
The slight disagreement at short $\tau$ may be due to measurement-induced qubit transitions that could be present even at low readout power~\cite{Slichter2012,Sank2016}.\\
\indent From our model with measurement errors taken into account, we extract $1/\sgam{00}=\nlb110\pm\nlb1~\us$, $1/\sgam{11}=\nlb77\pm\nlb1~\us$, $1/\sgam{10}=\nlb447\pm\nlb7~\us$, $1/\sgam{01}=\nlb400\pm\nlb5~\us$, 
$1/\ngam{10}=\nlb182\pm\nlb1~\us$, and
$1/\ngam{01}=\nlb6500\pm\nlb900~\us$.
Quoted parameter standard deviations reflect the uncertainty in the data, calculated using standard statistical techniques~\cite{Taylor1997}.
As a check of consistency, we calculate $T_1=\nlb(\sgam{10}+\ngam{10}+\sgam{01}+\ngam{01})^{-1}$, $\mathcal{P}_\mathrm{0}^\mathrm{eq}=\nlb(\sgam{10}+\ngam{10})T_1$, and $T_P\approx \nlb 2/(\sgam{00}+\sgam{11}+\sgam{10}+\sgam{01})$, and find that they agree with the independently measured values quoted above~\footnote{This approximate relation for $T_P$ is due to the fact that in the aforementioned charge-parity jump experiment~(Fig.~3) the qubit is taken out of equilibrium by the pulse sequence. The extracted $T_P$ approximately averages the conditional parity-switching rates corresponding to the qubit states $\ket{0}$ and $\ket{1}$.}.
A second transmon was found to have similar rates~\cite{SOM}.\\
\indent These rates have implications for our understanding of nonequilibrium QPs in our transmon qubits.
First, the limit on $T_1$ of this sample imposed by QPs is $(\sgam{10}+\sgam{01})^{-1}=\nlb211\pm 3~\us$, compared to a limit of $(\ngam{10}+\ngam{01})^{-1}=177\pm 2~\us$ imposed by all other loss mechanisms. This puts QP-induced dissipation on par with the sum of all other dissipation channels, contributing significantly to qubit relaxation~$\sgam{10}/\Gamma_{10}=\nlb0.29\pm\nlb0.01$.
Second, the ratio $\sgam{01}/\Gamma_{01}=\nlb0.94\pm 0.02$ indicates that QP-induced excitation accounts for the vast majority of the residual transmon excited-state population~[Fig. 4(a)], confirming previous suspicions~\cite{Wenner2013,Jin2015}.
Finally, $\sgam{01}/\sgam{10}=\nlb1.12\pm0.02$, 
which is direct evidence of a highly-energetic distribution of QPs.
Naïvely applying Fermi-Dirac statistics and detailed balance yields $\sgam{01}/\sgam{10}=\nlb\mathrm{exp}(-hf_{01}/k_\mathrm{B}T_\mathrm{eff}^\mathrm{qp})$, which predicts a negative effective QP temperature $T_\mathrm{eff}^\mathrm{qp} \approx-2~\mathrm{K}$ in our device. 
This is evidence that the QP energy distribution is not localized near the gap edge, but has a characteristic energy greater than $\Delta+hf_{01}$.
Conversely, $\ngam{01}/\ngam{10}=\nlb0.03\pm0.01$, indicating that the non-QP dissipative baths coupled to the transmon are relatively ``cold''~[Fig. 4(b)], with an effective temperature~$\sim 60~\mK$.
The observation that~$\sgam{11}>\nlb\sgam{00}$ is not yet explained by theoretical predictions~\cite{Catelani2014}.
We note that some weak dependence of QP dynamics on $\Ej/\Ec$ is expected, and following Appendix A of Ref.~\cite{Catelani2014} we find that the QP induced transition rates vary by less than a factor of 2 in the range $23<\Ej/\Ec<100$, with lower $\Ej/\Ec$ corresponding to increased QP sensitivity. To first order in perturbation theory, the ratio $\sgam{01}/\sgam{10}$ will not depend on $\Ej/\Ec$.\\
\begin{figure}
	\centering
	\includegraphics[width=\columnwidth]{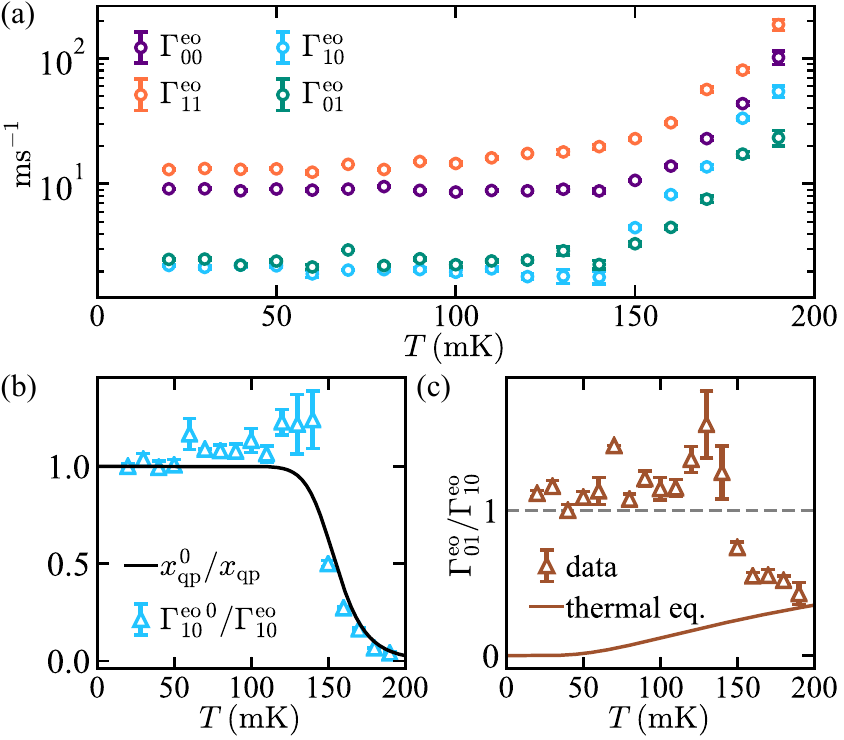} 
	\caption{Temperature dependence of qubit-state-conditioned parity-switching rates. 
    (a) Above $\sim140~\mathrm{mK}$, all rates begin to increase, and $\sgam{01}/\sgam{10}\leq 1$ suggests that thermally generated QPs begin to outnumber nonequilibrium QPs.
    (b) $1/\sgam{10}$ normalized by its base-temperature value $1/{\sgam{10}}^0$, as a function of temperature.
    The solid black line is a fit to the thermal dependence of $\xqpn/x_\mathrm{qp}$, which gives \mbox{$\xqpn\approx 1\times10^{-7}$.}
    (c) $\sgam{01}/\sgam{10}$ compared to predictions from detailed balance, assuming QPs are thermalized with the cryostat. Grey dashed line indicates the value above which~\mbox{$T_\mathrm{eff}^\mathrm{qp}\leq0$.}
	}
\end{figure}
We repeated the correlation measurement~[Fig.~4] at various mixing-chamber temperatures~$T$~[Fig.~5]. 
We find that all parity-switching rates $\sgam{ij}$ increase after $\sim 140~\mK$, at which point $T_1$, $T_P$, and $\sgam{01}/\sgam{10}$ all begin to decrease. 
Modeling the temperature dependence of these rates requires some ansatz about the QP energy distribution, which is typically assumed to be localized near the gap edge~\cite{Martinis2009,Catelani2011}. 
While this assumption appears not to be valid for QPs in our system, we use it to compare our results with other reports of QP density~$x_\mathrm{qp}^0$ in superconducting circuits.
If we further assume that the populations of nonequilibrium QPs and equilibrium QPs~\cite{Catelani2011} are independent, the total~$x_\mathrm{qp}$ is the sum:
\begin{equation}
x_\mathrm{qp}=x_\mathrm{qp}^0+\sqrt{2\pi k_\mathrm{B} T/\Delta} e^{-\Delta/k_\mathrm{B}T}.
\end{equation}
Here $\Delta=205~\mu\mathrm{eV}$, consistent with DC measurements of similar films ($\Delta$ increases with reduction of Al thickness)~\cite{Chubov1969}. The QP-induced relaxation rate $\sgam{10}$ should scale linearly with $x_\mathrm{qp}$~\cite{Martinis2009,Catelani2011}. We see this approximate scaling in our data~[Fig.~5(b)] with a slight decrease in $\sgam{10}$ with increasing temperature that is not predicted by our simple model, but has been previously observed~\cite{Martinis2009}.
This model yields~$x_\mathrm{qp}^0\approx\nlb1\times10^{-7}$, which agrees with other recent experiments~\cite{Vool2014,Nsanzineza2014,Aumentado2004,Pop2014,Wang2014}.\\
\indent Thus, we have shown that QPs are more energetic than a Fermi-Dirac distribution accounting for their apparent density $x_\mathrm{qp}^0$ would suggest. Further quantitative analysis of the measured parity switching rates, together with modeling of QP dynamics in our Al films, could reveal the energy range of QP-generating excitations. 
Proper filtering of RF lines, light-tight shielding~\cite{Barends2011,Corcoles2011}, and well-thermalized components are now standard ingredients for reducing the QP density which were included in our measurement setup~\cite{SOM}.
One should note that the authors of Ref.~\cite{Riste2013} reported $T_P$ one order of magnitude greater than what we have presented, with one experimental difference being a Cu readout cavity instead of a superconducting Al cavity.\\
\indent In conclusion, the correlations between charge-parity switches and qubit transitions in an offset-charge-sensitive transmon indicate that QP-induced loss can be responsible for a significant fraction of dissipation in state-of-the-art superconducting qubits. 
Additionally, we confirm that hot QPs with a highly-excited energy distribution are responsible for the residual excited-state population at low temperature in our samples. 
The techniques described above, building upon Ref.~\cite{Riste2013}, provide a tool to distinguish the influences of various experimental factors on QP generation and assess QP-reduction techniques, such as induced Abrikosov vortices~\cite{Wang2014,Nsanzineza2014,Vool2014,Taupin2016} or galvanically connected QP traps~\cite{Booth1993,Court2008,Peltonen2011,Rajauria2012,VanWoerkom2015,Riwar2016,Hosseinkhani2017,Patel2017}.\\
\indent We acknowledge insightful discussions with \mbox{Gianluigi} \mbox{Catelani,} \mbox{Leo DiCarlo,} \mbox{Yvonne Gao,} \mbox{Leonid Glazman,} \mbox{Ioan Pop,} \mbox{Dan Prober,} \mbox{Rob Schoelkopf,} and \mbox{Uri Vool.} 
Facilities use was supported by YINQE, the Yale SEAS cleanroom, and NSF MRSEC DMR 1119826. 
This research was supported by ARO under Grant No. W911NF-14-1-0011, by MURI-ONR under Grant No. N00014-16-1-2270, and NSF DMR Grant No. 1603243.
GdL acknowledges support from the European Union’s Horizon 2020 research and innovation programme under the Marie Skłodowska-Curie grant agreement No. 656129. M. Houzet acknowledges support from the European Union's FP7 programme through the Marie-Skłodowska-Curie Grant Agreement 600382.

\bibliography{biblio_prl.bib}

\chapter{}

\clearpage
\widetext
\begin{center}
\textbf{\large Supplemental materials for ``Hot nonequilibrium quasiparticles in transmon qubits''}
\end{center}
\setcounter{equation}{0}
\setcounter{figure}{0}
\setcounter{table}{0}
\setcounter{page}{1}
\makeatletter
\renewcommand{\theequation}{S\arabic{equation}}
\renewcommand{\thefigure}{S\arabic{figure}}
\renewcommand{\thetable}{S\arabic{table}}

\renewcommand{\bibnumfmt}[1]{[S#1]}
\renewcommand{\citenumfont}[1]{S#1}

\section{SUMMARY OF DEVICES}
\begin{table}[h]
     \centering
     \caption{Summary of device parameters. } 
	\begin{tabular}{c|*{10}{c}}
    \hline
    \hline
	Sample & $\overline{f_{01}}$(GHz) & $2\delta f_{01}$(MHz)& $T_1$($\us$) & $T_P$($\us$) & $1/\sgam{00}$($\us$) & $1/\sgam{11}$($\us$) & $1/\sgam{10}$($\us$) & $1/\sgam{01}$($\us$) & $1/\ngam{10}$($\us$) & $1/\ngam{01}$($\us$)\\
	\hline
	A & 4.400 & 3.18 & 95 $\pm$ 5 & 77 $\pm$ 1 & 110 $\pm$ 1& 77 $\pm$ 1 & 447 $\pm$ 7 & 400 $\pm$ 5 & 182 $\pm$ 1& 6500 $\pm$ 900\\
	B & 4.255 & 4.96 & 44 $\pm$ 2 & 96 $\pm$ 1  & 135 $\pm$ 2& 92 $\pm$ 2 & 920 $\pm$ 80 & 400 $\pm$ 10 & 61 $\pm$ 1& 10000 $\pm$ 4000\\
    \hline
    \hline
	\end{tabular}
\end{table}
Sample A is the device reported in the main text.
Sample B was measured in a separate cooldown under nominally identical conditions, though we did not study the temperature dependence of its conditional transition rates.
Error estimates on all parameters are extracted from experimental uncertainty and do not reflect slow changes in those quantities over time. Reported $T_1$ values are obtained from a free-decay measurement independent of the charge-parity-correlation experiment used to extract QP-induced transition rates, and match those results within experimental fluctuations.
Combinations of these rates give intuitive metrics by which to assess QP loss in our transmons. 
The limit on $T_1$ of sample A (sample B) imposed by QPs is $(\sgam{10}+\sgam{01})^{-1}=\nlb211\pm 3~\us\,(278\pm 8~\us)$, while all non-QP loss mechanisms limit~$T_1$ to~$(\ngam{10}+\ngam{01})^{-1}=177\pm 2~\us\,(61\pm 1~\us)$. QP-induced transitions account for~$\sgam{10}/\Gamma_{10}=\nlb0.29\pm\nlb0.01\,(0.06\pm 0.01)$ of all relaxation events, and~$\sgam{01}/\Gamma_{01}=\nlb0.94\pm 0.02\,(0.96\pm 0.03)$ of all excitation events.
The ratio of QP-induced excitation and relaxation rates $\sgam{01}/\sgam{10}=\nlb1.12\pm0.02\,(2.3\pm 0.2)$, 
indicates that the QPs are ``hot".
Conversely, we find that other the combination of all other dissipative baths are ``cold": $\ngam{01}/\ngam{10}=\nlb0.03\pm0.01\,(0.006\pm 0.002)$.\\
\section{EXPERIMENTAL SETUP}
\indent Samples were mounted in a superconducting Al 3D rectangular readout cavity with resonant frequency frequency~$f_\mathrm{r}=\nlb9.204~\GHz$ and linewidth~$\kappa_\mathrm{r}/2\pi=\nlb1.8~\MHz$.
These devices were measured in the dispersive regime of circuit-QED~\cite{Blais2004} (dispersive shift~$\chi_\mathrm{qr}/2\pi=\nlb3.8~\MHz$), and a Josephson Parametric Converter (JPC)~\cite{Bergeal2010} was used to achieve a single-shot qubit-readout fidelity of~$\approx\nlb0.97$ in $3.84~\us$ with an average readout-resonator occupation $\bar{n}\approx\nlb 3$.\\
\indent QP dynamics may be influenced by various aspects of the experimental setup including RF filtering, radiation shielding, use of magnetic materials, and thermalization of the sample. 
Fig. S1 shows a schematic representation of the RF-lines and shielding inside the cryostat. 
Magnetic fields can induce vortices which have been shown to decrease QP loss, though this advantage can be undermined by vortex flow dissipation if the magnetic field at the sample is too strong. 
The transmon was mounted in a separate Cryoperm magnetic shield from the JPC, and special care was taken to not include any strongly magnetic materials inside the shield in order to establish a baseline understanding of QP dynamics in our system. 
The aluminum sample holder/readout cavity was mounted to a copper bracket using brass screws and molybdenum washers, which were tested prior to use with a magnetometer.
A copper plate coated with carbon black suspended in Stycast was placed inside the Cryoperm shield and thermalized to the sample mounting bracket with copper braid. 
This is an attempt to absorb any photons that leak into the shield. 
A copper thermalization braid was attached directly to the Al readout cavity, providing a direct thermal link to the mixing chamber stage. 
\begin{figure*}[t]
	\centering
	\includegraphics[width=\textwidth]{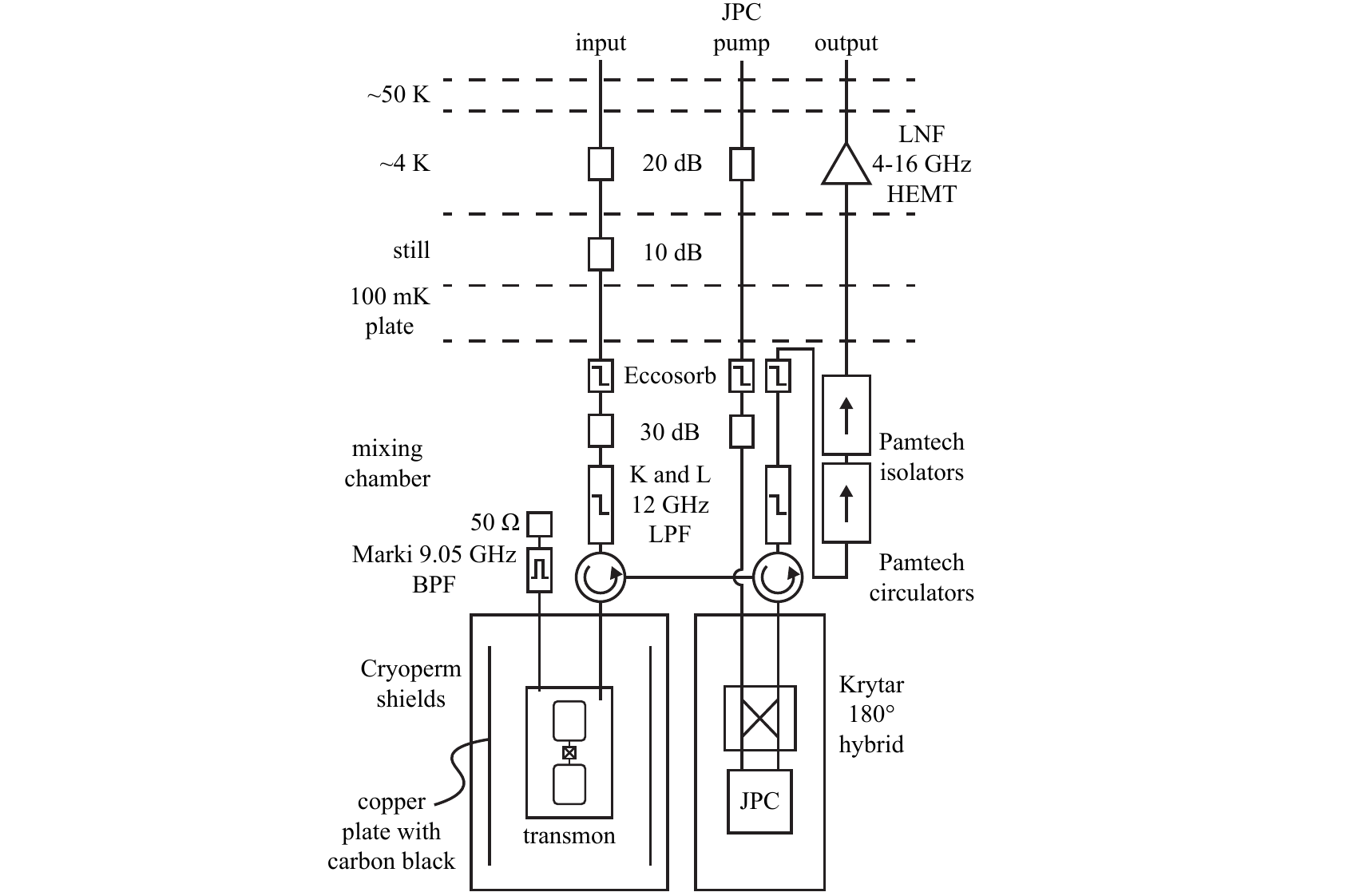} 
	\caption[width=\textwidth]{Outline of radio-frequency components in our experimental setup.
    }   
\end{figure*}
\section{DEVICE FABRICATION}
The devices were patterned in a bilayer of Microposit A4 PMMA and Microposit EL13 PMMA-MAA copolymer on a c-plane sapphire substrate by a $100~\mathrm{keV}$ Vistec EBPG $5000+$ using standard electron-beam lithography techniques. The JJ mask was designed using the ``bridge-free-technique"~\cite{Lecocq2011}.
The JJ electrodes were formed from $20$ and $30~\nm$ thin-film Al, e-beam evaporated in a Plassys UMS300 at an angle of $\pm 20^\circ$, respectively.
\section{PSD OF CHARGE-PARITY SWITCHES}
Repeated measurements of charge-parity produce a parity-jump trace that looks like a symmetric random telegraph signal with variance~$=1$.
The power spectral density of these parity fluctuations, $S_{PP}$, is fit to a modified Lorentzian of the form
\begin{equation}
S_{PP}(f)=\frac{4\F^2/T_P}{(2/T_P)^2+(2\pi f)^2}+(1-\mathcal{F}^2)\Delta t_\mathrm{exp}
\end{equation}
Above, $T_P=77\pm1~\us$ is the characteristic charge-parity switching rate, $\F=0.91\pm0.01$ is the fidelity of the parity mapping, and $\Delta t_\mathrm{exp}=10~\us$ is the sampling period of the signal. This model assumes that the detection errors leading to non-unity $\F$ are uncorrelated with charge-parity, though $T_1$-errors tend to bias toward measuring even charge parity (discussed below). A``chi-squared" analysis of the model suggests that this has a negligible effect on the output of the model. For more details, see Ref.~\cite{Riste2013}.
\section{MODELING CORRELATIONS BETWEEN QUBIT TRANSITIONS AND CHARGE-PARITY SWITCHES}
We measured correlations between charge-parity switches and qubit transitions, which reveals the extent to which the qubit coherence is limited by nonequilibrium quasiparticle excitations.
To correlate these processes, we perform two charge-parity mapping sequences, separated by a variable delay $\tau$~[Fig. S2]. From this, we sort our measurement sequences conditioning on starting in qubit state $i$ and parity $p$, and ending up in qubit state $j$ and parity $p'$. We compute two quantities from this data: the conditioned probabilities of all of these events $\tilde{\rho}(j,pp'|i)(\tau)$, and the qubit-state-conditioned charge-parity autocorrelation function~$\pp{ij}(\tau)$.
To model the dynamics between states of the system, we define a master equation describing the dynamics of joint qubit-state and charge-parity occupation probabilities $\rho_i^\alpha$.
\begin{equation}
\begin{split}
\dot{\rho_i^\alpha}=&-(\Gamma_{i\bar{i}}^{\alpha\bar{\alpha}}+\Gamma_{ii}^{\alpha\bar{\alpha}}+\Gamma_{i\bar{i}}^{\alpha\alpha})\rho_i^\alpha \\&+ \Gamma_{\bar{i}i}^{\bar{\alpha}\alpha} \rho_{\bar{i}}^{\bar{\alpha}}+\Gamma_{ii}^{\bar{\alpha}\alpha} \rho_{i}^{\bar{\alpha}}
+\Gamma_{\bar{i}i}^{\alpha\alpha} \rho_{\bar{i}}^{\alpha}.
\end{split}
\end{equation}
Here, $\Gamma_{i\bar{i}}^{\alpha\bar{\alpha}}$ is a conditional transition rate, with $i$ ($\bar{i}$) and $\alpha$ ($\bar\alpha$) denoting the conditioned (other) qubit state and charge parity, respectively.
Because the charge dispersion of the transmon energy levels is small relative to the scale of thermal fluctuations, the conditional rates are symmetric with the exchange of $\alpha$ and $\bar{\alpha}$.
We evolve this master equation with initial conditions set by conditioning on the initial qubit and charge-parity state.
The full model is solved numerically and fit to measured values of all eight permutations of $\tilde{\rho}(j,pp'|i)(\tau)$ and all four permutations of $\pp{ij}(\tau)$, to extract $\Gamma_{00}^{\alpha\bar{\alpha}}$, $\Gamma_{11}^{\alpha\bar{\alpha}}$, $\Gamma_{10}^{\alpha\bar{\alpha}}$, $\Gamma_{01}^{\alpha\bar{\alpha}}$, $\Gamma_{10}^{\alpha\alpha}$, and $\Gamma_{01}^{\alpha\alpha}$.\\
\indent The measured values of $\tilde{\rho}(j,pp'|i)(\tau)$ and $\pp{ij}(\tau)$ are susceptible to various measurement infidelities that are not included in the model above, and we must modify our fit functions to include these infidelities.
Single state-discrimination errors will on average decrease $\pp{ij}(\tau)$, and $T_1$ errors during the parity mapping will impart an infidelity that depends on both the charge-parity and the qubit state at the start of the parity mapping.
Other measurement inefficiencies are approximately independent of qubit state and charge parity, which contribute to a global fidelity $\F_g$ of the parity-mapping sequence.
For example, because $\n$ varies uncontrollably in time, each sequence of pulse calibrations and parity-autocorrelation measurement must be completed on a timescale faster than a few minutes.
Any variation of $\n$ between the tuning of pulses and the completion of the experiment will introduce qubit-pulse errors, which along with qubit dephasing during $\tng$, contribute to $\F_g$. 
In practice, $\F_g$ is occasionally very low, which we attribute to spontaneous jumps in $\n$ between the the time when $\df$ is determined and the correlation measurement.
Since we do not know $\F_g$ \emph{a priori}, we include it in the model as an additional fit parameter, and exclude independent measurement sequences which fall below a threshold $\F_g$. 
This threshold is 0.5 at low temperatures, where the vast majority of measurements meet this criteria. This threshold must be relaxed at higher fridge temperatures due to increased qubit dephasing. \\
\indent State-discrimination errors can be sufficiently reduced by ignoring measurement sequences in which any of the four measurements do not meet a stringent state-assignment threshold.
We histogram all qubit measurements, fit to a sum of two Gaussian distributions, and exclude measurement sequences where any of the four measurements fall near the half-way point between distributions.
In practice, this thresholding removes between $10\%$ and $50\%$ of measurement sequences, depending on the amplitude and integration time of the readout signal, in order to achieve state-discrimination fidelity of greater than $0.9999$. The readout amplitude was limited to an average photon number $\bar{n}\approx3$ to avoid measurement induced qubit transitions~\cite{Slichter2012,Sank2016}. \\
\indent Each parity-mapping sequence consists of an initial qubit measurement, the Ramsey pulses for parity-mapping, and a final qubit measurement.
Because of stringent thresholding, we assume state-assignment with perfect fidelity that is achieved at the midpoint of the readout pulse.
There is therefore a time $\tau_1$ between the midpoint of $M_1$ and the beginning of the Ramsey pulses, and time $\tau_2$ between the end of the Ramsey pulses and the midpoint of $M_2$, during which $T_1$ errors can occur~[Fig.~S2]. Errors during $\tau_1$ and $\tau_2$ from $T_1$ events are included explicitly in the model, and errors between the $\pi/2$-pulses are included implicitly via a global mapping fidelity $\F_g$.\\
\begin{figure*}
	\centering
	\includegraphics[width=\textwidth]{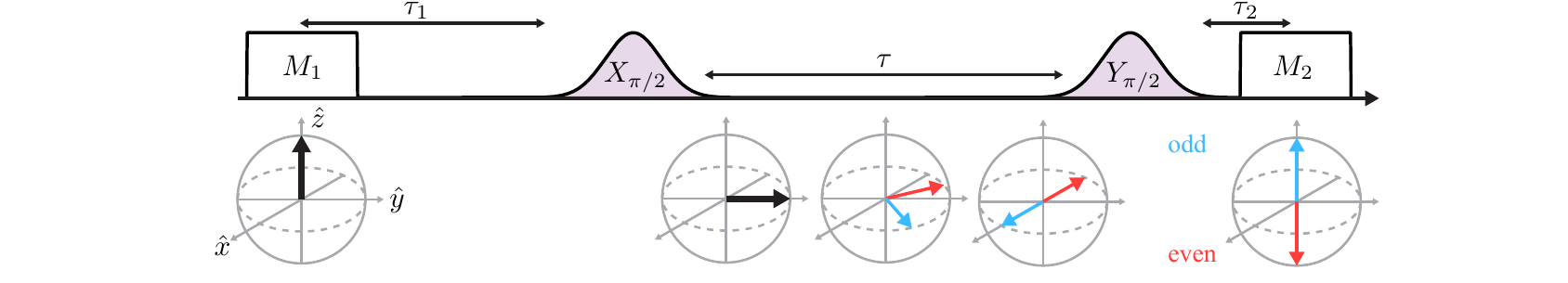} 
	\caption[width=\textwidth]{Charge-parity mapping pulse sequence (not to scale). The charge-parity is defined as $P= (2M_1-1)(2M_2-1)$. The first $\pi/2$-pulse brings both charge-parity Bloch vectors to the equator. After a precise delay $\tau = 1/4\df$, the Bloch vectors are anti-parallel. A second $\pi/2$-pulse completes the operation, enacting an effective $\pi$-pulse conditioned on being in the even charge-parity state, regardless of the outcome $m_1$. We can change the conditioning of the mapping by changing the phase of the first $\pi/2$-pulse by $180^\circ$. 
    }   
\end{figure*}
\indent Qubit-state dependent $T_1$ events affect the fidelity with which we determine the charge-parity. 
For example, let's say the parity-mapping sequence is chosen such that it enacts a $\pi$-pulse conditioned on being in the even charge-parity state (this will vary in the following discussion). If the system is in state $\ket{0\mathrm{,odd}}$, one would expect to measure $m_1=0\rightarrow m_2=0$, but $T_1$ errors will appear as $0\rightarrow 1$ with a probability $\Gamma_{01}(\tau_1+\tau_2)$.
If the system state is $\ket{0\mathrm{,even}}$, one would expect to measure $0\rightarrow 1$, but $T_1$ errors will appear as $0\rightarrow 0$ with a probability $(\Gamma_{01}\tau_1+\Gamma_{10}\tau_2)$.
Similar expressions can be found for the system starting in $\ket{1}$.
Since there is no physical preference for even or odd parity, we average over parity dependence in the error rates and only consider the probability of starting in an initial state.
However, parity-dependent errors will introduce artificial correlations between $P$ and $P'$. 
To remedy this, we vary whether each parity-mapping sequence performs an effective $\pi$-pulse on the even- or odd-charge-parity state.
Assuming near-perfect state discrimination fidelity and equal probability to measure odd or even parity (with balanced pulse conditioning), these errors will only depend on the qubit state at the beginning of the mapping. For the first parity-mapping sequence $P$, we define an error probability 
\begin{equation}
\gamma_P^{ij}= (1-\F_g)+(\mathcal{P}_0^{ij}(\tau_1\Gamma_{01}+\tau_2\Gamma_{10})+\mathcal{P}_1^{ij}(\tau_1+\tau_2)\Gamma_{10})/2
\end{equation}
Above, $\mathcal{P}_n^{ij}$ is the probability that $m_1=n$ at the beginning of $P$ in measurement sequences with qubit-conditioning $m_2=i$ and $m_3=j$. Similarly, for the second parity mapping sequence $P'$ we define
\begin{equation}
\gamma_{P'}^{j}=(1-\F_g)+((\tau_1+\tau_2)\Gamma_{j\overline{j}}+(\tau_1\Gamma_{j\overline{j}}+\tau_2\Gamma_{\overline{j} j}))/2
\end{equation}
This error probability is independent of $i$, and does not have additional qubit-state weighting because we assume near-perfect conditioning of $j$.\\
\indent Without accounting for any errors, $\tilde{\rho}(j,pp'|i)(\tau)=\rho_i^{p}(0)\rho_j^{p'}(\tau)$. Errors in the determination of $\rho_i^{p}(0)$ shuffle the initial probability from the conditioned parity $\rho_i^{p}(0)$ to the other parity $\rho_i^{\overline{p}}(0)$ with a rate $\gamma_P^{ij}$. We evolve the master equation with these errors accounted for in the initial conditions, in that the conditioned probability~$\rho_i^{\overline{p}}(0)$ is no longer unity. Then, applying errors in the second parity mapping explicitly, we find:
\begin{equation}
\tilde{\rho}(j,pp'|i)(\tau)=(1-\gamma_{P'}^{j})\rho_j^{p'}(\tau)+\gamma_{P'}^{j}\rho_j^{\overline{p'}}(\tau).
\end{equation}
We calculate~$\pp{ij}(\tau)$ directly from these conditional probabilities
\begin{equation}
\pp{ij}(\tau)=\frac{\tilde{\rho}(j,+1|i)(\tau)-\tilde{\rho}(j,-1|i)(\tau)}{\tilde{\rho}(j,+1|i)(\tau)+\tilde{\rho}(j,-1|i)(\tau)}.
\end{equation}
To extract the rates quoted in Table S1, we fit to all eight permutations of~$\tilde{\rho}(j,pp'|i)(\tau)$ and all four permutations of~$\pp{ij}(\tau)$ simultaneously.\\
\indent Our analysis relies on the above model to accurately extract qubit-state-conditioned QP tunneling rates, and we claim that the ratio $\sgam{01}/\sgam{10}$ is well-captured by the model. To illustrate this, we plot our data along with predicted curves $\pp{ij}(\tau)$ for various fixed $\sgam{01}/\sgam{10}$~[Fig. S3]. This model is constructed by first fixing $\sgam{00}$ and $\sgam{11}$ to the values extracted from the fit to the data. These rates approximately fix $T_P$ to the value extracted in the main text. Then, we adjust $\sgam{10}$, $\sgam{01}$, $\ngam{10}$, and $\ngam{01}$ under the constraint that $T_1$, $T_P$, and $\mathcal{P}_\mathrm{0}^\mathrm{eq}$ are fixed to their independently measured values for all chosen values of $\sgam{01}/\sgam{10}$. As displayed in Fig. S3, the model qualitatively deviates from the data when $\sgam{01}/\sgam{10}$ is less than $\approx 1$.
\begin{figure*}
	\centering
	\includegraphics[width=\textwidth]{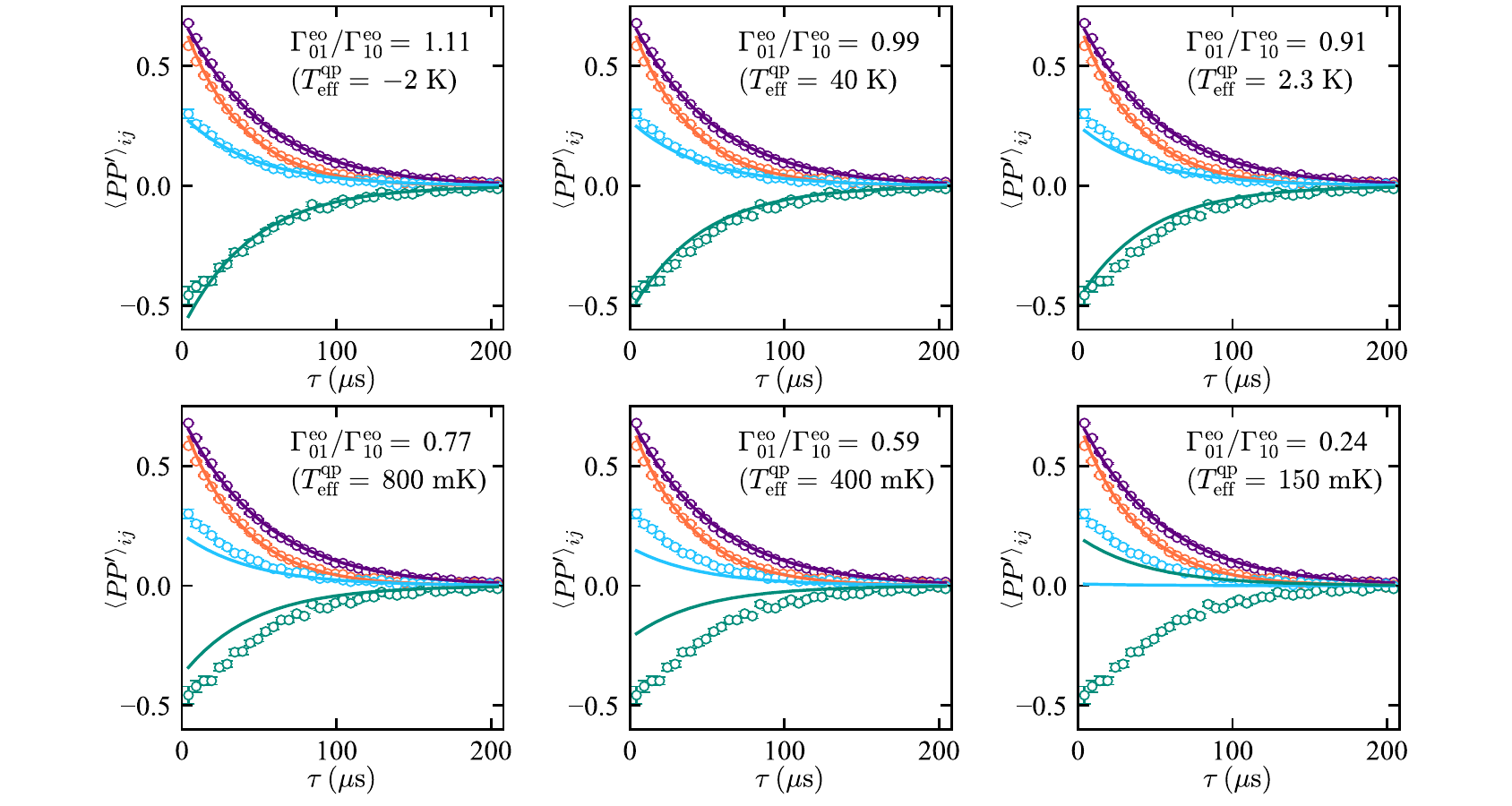} 
	\caption[width=\textwidth]{Charge-parity-autocorrelation data [Fig. 4] overlaid with the master equation model of $\pp{ij}(\tau)$ for varying $\sgam{01}/\sgam{10}$. Deviation from the data when setting $\sgam{01}/\sgam{10}<1$ shows that the model is sensitive to small fluctuations in fit parameters. Fixing the effective QP temperature to account for the apparent density of nonequilibrium QPs ($T_\mathrm{eff}^\mathrm{qp}=150~\mK$ for $x_\mathrm{qp}\approx 10^{-7}$) clearly does not accurately describe their energy distribution.
    }   
\end{figure*}
 
\bibliography{biblio_prl.bib}

\end{document}